# Electron Recoil via Sample Momentum Transfer in Optical-Mode Excitation

Akira Yasuhara[1], Yamato Kirii[2], Takumi Sannomiya[2,*]

[1]JEOL Ltd., 3-1-2 Musashino, Akishima, Tokyo, 196-8558, Japan.
[2] Department of Materials Science and Engineering, School of Materials and Chemical Technologies, Institute of Science Tokyo, 4259 Nagatsuta, Midoriku, Yokohama, 226-8503, Japan

**ABSTRACT**.
The interaction between free electrons and optical modes underlies a variety of quantum and nanoscale light–matter phenomena, yet the associated momentum exchange with the sample largely remained overlooked. Here, we experimentally demonstrate the momentum transfer from free electrons to planar samples during optical mode excitation using momentum-resolved electron energy-loss spectroscopy. The momentum transfer to the sample modifies the apparent dispersion relation which is significant when the planner sample is tilted. Under specific conditions, the sample receives momentum opposite to the electron beam direction.

Coherent optical excitation by free electrons has attracted renewed interest for its potential to reveal quantum correlations and entanglement between electrons and photons or quasiparticles generated through the electromagnetic interaction [1–5]. Such processes, governed by energy or/and momentum conservation, form the basis for a range of emerging concepts in free-electron quantum optics, including quantum state generation [6–8], quantum measurement [9], or signal enhancement [10,11]. To excite optical modes and emit photons or quasiparticles, an electron typically interacts with a material sample that mediates the electromagnetic interaction and imposed energy–momentum exchange. In such interactions, part of the electron's momentum may be transferred to the sample to satisfy momentum conservation [12–14]. In contrast to the energy conservation, which can be probed by spectral selection of correlated photons and electrons, the momentum correlation have been investigated using planar samples to minimize the complexity arising from in-plane momentum transfer to nanostructured materials [14,15]. Nevertheless, even in planar geometries, momentum transfer perpendicular to the surface occurs to ensure total momentum conservation. However, quantitative manifestation of momentum transfer to the sample through electron recoil during optical excitation has not yet been experimentally investigated.

Here, we show how such momentum transfer occurs in planar samples through the excitation of in-plane optical modes by tilting the samples and how it modifies the apparent dispersion relation using the momentum-resolved electron energy-loss spectroscopy (qEELS) based on transmission electron microscopy (TEM) [16–18]. We further demonstrate that under specific conditions, the optical excitation can induce a recoil effect in which the sample effectively receives momentum directed opposite to the incident electron beam.

For this demonstration, we prepared a specimen consisting of metal films deposited on an amorphous silicon nitride ($Si_3N_4$) membrane of 180 nm thick, as shown in Fig. 1a. The membrane has an array of holes with a periodicity of 200 nm, which works a measure for the sample tilt and the scale calibration as we discuss later. (Fig. 1b) The sample was tilted by angle $\varphi$ around the $y$ axis, as indicated in the figure. In this configuration, surface plasmon polaritons (SPPs) on both sides of the aluminum (Al) films are dominantly excited. (see Supplemental Material (SM) for the details).

To evaluate the DRs and the momentum transfer from the electron to the sample, low-angle diffraction patterns and qEELS spectra are obtained at different sample tilt angles, as shown in Fig.2a. The measurement was all performed at 200 kV acceleration using a JEM-F200 (JEOL Ltd., Japan) instrument equipped with a GIF Continuum ER (Gatan Inc., USA). The electron beam was expanded ~10 μm to achieve sufficient momentum resolution and a slit mask is used for the qEELS spectral acquisition. (More experimental details can be found in the previous study [17].) The observed diffraction spots reflect the periodicity of the holes. Without the sample tilt (Fig.2, middle $\varphi = 0°$), where the diffraction pattern

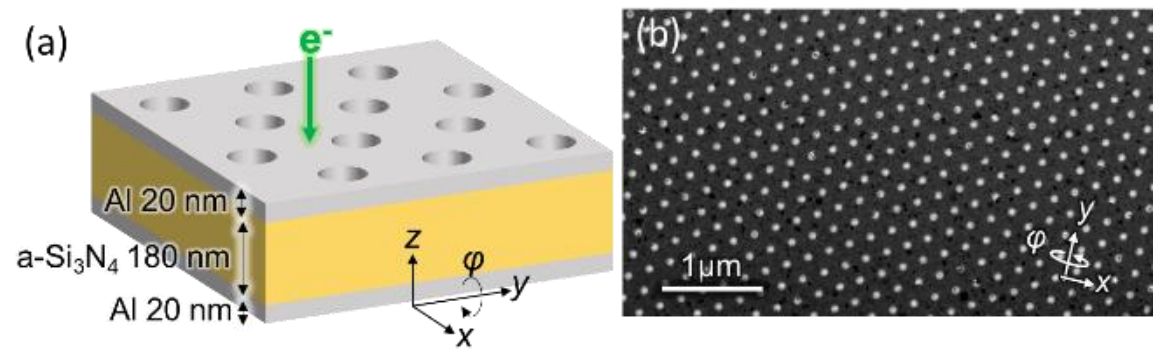

FIG 1. (a) Illustration of the sample and coordinate. (b) TEM bright field image of the sample.

*Contact author: sannomiya.t.aa@m.titech.ac.jp

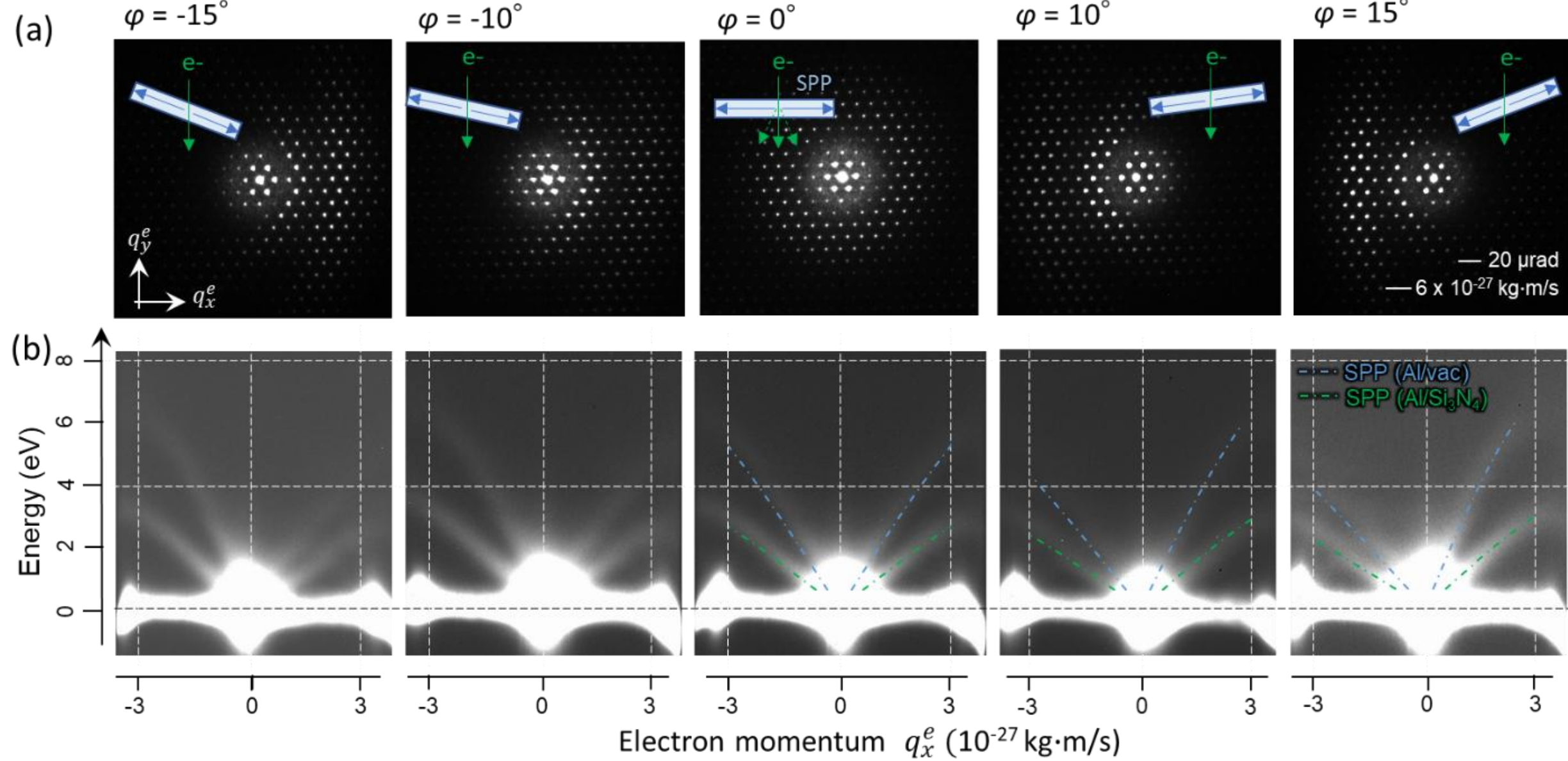

FIG 2. (a) Elastic electron diffraction patterns with different sample tilt angles $\varphi = 0, \pm 10, \pm 15$ ° around the $y$ axis. (b) DRs obtained by qEELS at corresponding sample tilt angles. Calculated DR lines of SPPs at the vacuum and $Si_3N_4$ interfaces are overlaid for $\varphi = 0, 10, 15$ °.

intensity is concentrically distributed, the DR curves of the electrons in Fig.2b nicely matches the expected DR curves of single interface SPPs, which are superposed on the figure. (see also SM for the existent modes [19]) The SPP DRs are distributed symmetrically on both positive and negative sides of the horizontal momentum axis. The diffraction spot intensity distributions due to the periodic holes in Fig. 2 work as measures for the sample tilt alignment as well as for the momentum calibration. The optical dispersion overlap from the periodicity is not clearly observed because of the energy resolution and the broad features of the SPP. (see SM [19])

By tilting the sample, the diffraction pattern intensities in Fig. 2a become asymmetric along the $q_x$ direction with non-concentric dark ring patterns similar to Laue zones in 3D crystals. [20,21] In the qEELS results in Fig.2b, the observed electron DR looks also asymmetric by the sample tilt, showing "inclined" DR lines, which seemingly not directly correspond to the pure SPP DRs. We note that the overlap of the dispersion line stemming from the periodic lattice is slightly visible on the negative (positive) $q_x^{\mathrm{e}}$ side at $\varphi$ = +(-)15°. While such modification of the electron dispersion by SPP excitation with sample tilts has previously been shown, [22–24] the momentum transfer to the sample had not been discussed. We here elucidate this dispersion modification in terms of the momentum transfer to the sample, which is schematically illustrated in Fig.3. (see also SM for the case including photons) For the elastic scattering, one can only consider the momentum transfer to the lattice whereas for the inelastic scattering optical modes (SPPs) should be considered (Fig.3a). Because the sample is thin, the reciprocal lattice is elongated in the $z$ direction, also forming the fringe patterns in the $z$ direction by the finite sample thickness (Fig. 3b, see also SM). [19,25] For the elastic scattering, intersections of the reciprocal lattice and the Ewald sphere (red lines in Fig.3b) corresponds to the momentum conservation. Due to the finite size of the hole forming non-uniform reciprocal lattice intensities and to the curvature of the Ewald sphere going across the $z$-fringe node positions, forming the Laue zone-like dark rings in the diffraction patterns found in Fig.2a.

For the inelastic scattering, where the length of the electron momentum vector changes from $q_0^{\mathrm{e}}$ to $q_\pm^{\mathrm{e}}$, the momentum conservation of the entire system including optical modes and the sample should be considered, as show in Fig.3c. Since we handle the scattering of the energy-loss electron, the Ewald sphere is drawn with the radius of $q_\pm^{\mathrm{e}}$(Fig.3c). Here we consider weak interaction where only a single particle (SPP, or optical mode) is generated by one electron as well as no kinetic energy of the sample, which is reasonable for this system. [14,26] The momentum vectors of the scattered electron (e.g. $q_+^{\mathrm{e}}$), optical mode ($q_+^{\mathrm{p}}$), and sample ($q_+^{\mathrm{s}}$) are added to meet the initial momentum of the electron ($q_0^{\mathrm{e}}$). Since the optical mode (SPP) is supported by the thin sample, this mode can be described as a cylinder elongated in the $z$-direction in the reciprocal (momentum) space, as described as diffuse violet lines in Fig.3c. (see also SM for the 3D representation) [19] Without the sample tilt ($\varphi = 0°$), this scattering process is symmetric around the $z$ axis.

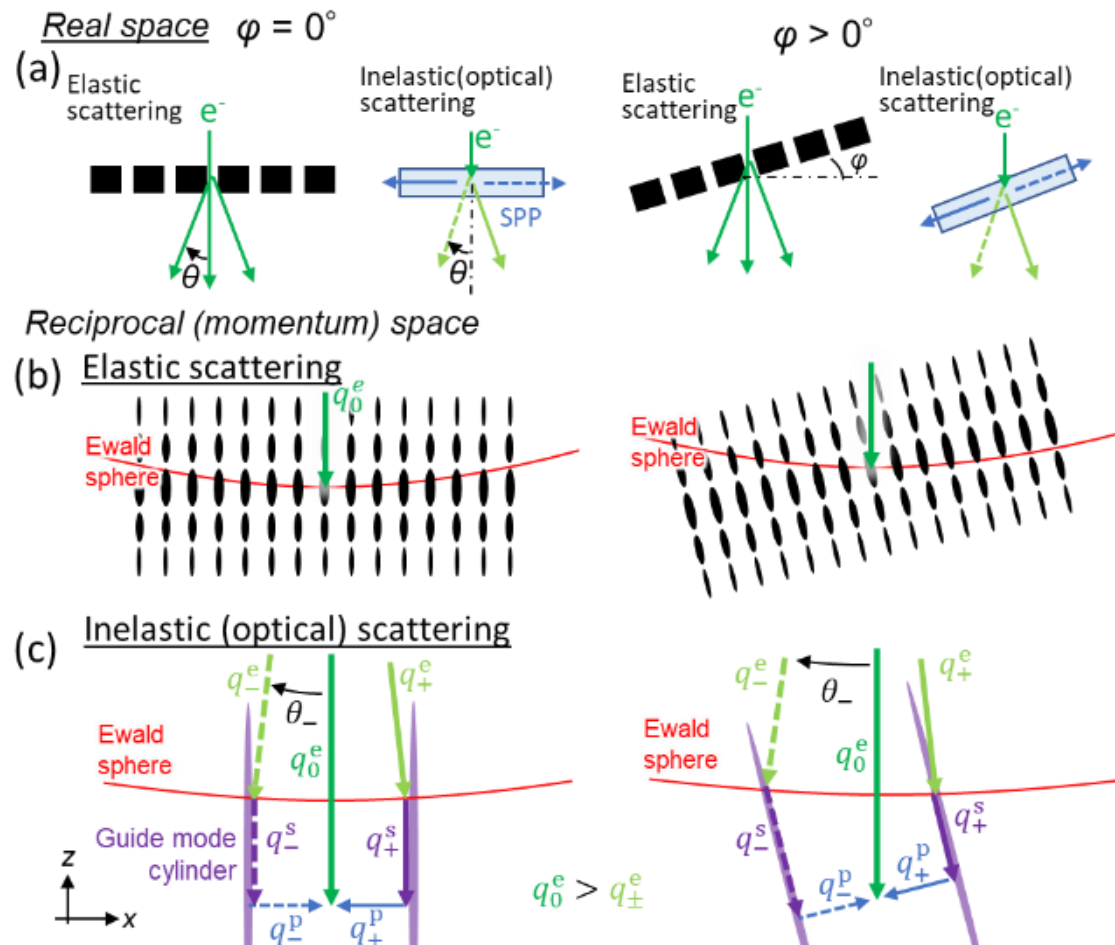

FIG 3. (a) Illustrations of the diffraction (left) without and (right) with sample tilts. (b) Reciprocal space describing the elastic diffraction by the periodic holes structures in the membrane. (c) Reciprocal space describing the inelastic scattering due to optical mode on the film plane. The momentum $q$ is expressed with the superscript of e, p, and s respectively standing for the electron, SPP and sample. The subscript sign indicates the same series of the inelastic scattering process. $q_0^{\mathrm{e}}$ is the incident electron momentum.

By tilting the sample ($\varphi > 0$), the reciprocal lattice points related to the periodic holes, as well as the optical mode cylinder for the inelastic scattering, are inclined, as described in the right panels of Fig.3. For the elastic scattering, the sample tilt results in the asymmetric intersections of the Ewald sphere, which explains the asymmetric intensity distributions of Fig.2a in the tilted conditions. Also in the inelastic scattering process by optical excitation, the inclined optical mode cylinder results in the asymmetric intersections with the Ewald sphere, as described in the right panel of Fig.3c. This asymmetry exerts a larger in-plane momentum of electrons in the negative side ($q_-^{\mathrm{e}}$) and a smaller in the positive side ($q_+^{\mathrm{e}}$), which induces apparent "inclined" DR of electrons, as in the experiment of Fig.2b.

The cross-section of the Ewald sphere and the optical mode cylinder at a given energy (Fig. 3c) can also be experimentally confirmed by energy-filtered diffraction patterns, as shown in Fig. 4 (obtained in a planer area without lattice structure to avoid dispersion cone overlaps). The dispersion circles, which is concentric without the sample tilt (Fig.4a), become distorted and displaced from the center by tilting the sample (Fig.4b). With the sample tilt in Fig. 4b, the intensity looks weaker on the left side because the dispersion curve is more blurred by inclination and the reciprocal intensity is weaker for larger $q_-^{\mathrm{s}}$ on the negative $q_x^{\mathrm{e}}$ side at $\varphi = +15°$ in Fig. 2b.

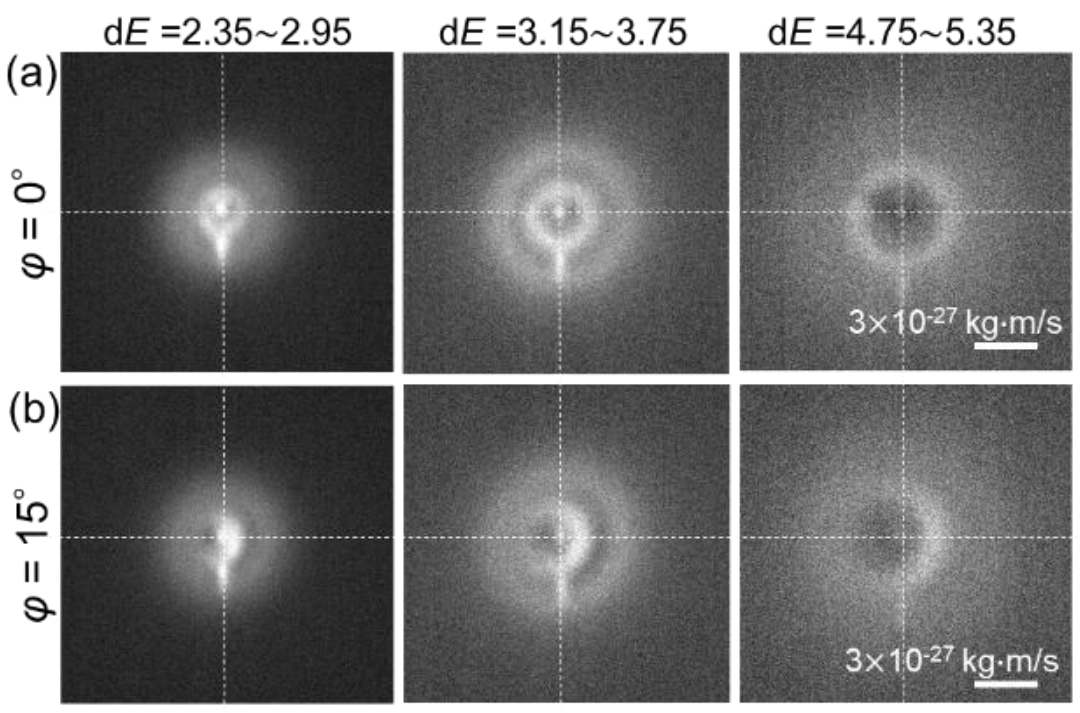

FIG 4. Energy-filtered diffraction patterns (a) witout and (b) with sample tilts, obtained at a planer position without lattice structures.

This scattering process can be more quantitatively described using the momenta of the electron $q^{\mathrm{e}}$, optical mode $q^{\mathrm{p}}$, and sample $q^{\mathrm{s}}$ after scattering with the sampled tilt $\varphi > 0$. Here we only consider the deflection in the $q_-^{\mathrm{e}}$ side with the electron deflection angle $\theta_-$ which is enhanced by launching the optical mode (left-hand side of each illustration in Fig. 3c). The opposite scattering side can be obtained by flipping the sign of $\varphi$. The momentum conservations are expressed as $q_-^{\mathrm{e}}\sin\theta_- = q_-^{\mathrm{s}}\sin\varphi + q_-^{\mathrm{p}}\cos\varphi$ for the $x$ direction, and $q_0^{\mathrm{e}} = q_-^{\mathrm{e}}\cos\theta_- + q_-^{\mathrm{s}}\cos\varphi - q_-^{\mathrm{p}}\sin\varphi$ for the $z$ direction. $q_0^{\mathrm{e}}$ is the initial momentum of the incident electron. When the DR of the optical mode is known, as in Fig.1, the electron scattering angle $\theta_-$ and the transferred momentum to the sample $q_-^{\mathrm{s}}$ can be calculated. For sufficiently small $\theta_-$ (<< 1), which is the case for optical excitations, $\theta_-$ and $q^{\mathrm{s}}$ are expressed as:

$$\theta_- = \frac{1}{q_-^{\mathrm{e}}}\{q_-^{\mathrm{p}}\cos\varphi + (\Delta q^{\mathrm{e}} + q_-^{\mathrm{p}}\sin\varphi)\tan\varphi\}, \quad (1)$$

$$q_-^{\mathrm{s}} = \frac{1}{\cos\varphi}(\Delta q^{\mathrm{e}} + q_-^{\mathrm{p}}\sin\varphi). \quad (2)$$

Here $\Delta q^{\mathrm{e}} = q_0^{\mathrm{e}} - q_\pm^{\mathrm{e}}$ is the momentum amplitude change of the electron by inelastic scattering. We note that $q^{\mathrm{s}}$ vector is always perpendicular to the sample surface.

The calculated electron DRs, corresponding to the $x$ component of the electron momentum $q_x^{\mathrm{e}} = q_\pm^{\mathrm{s}}\sin\theta_\pm$ at a given loss energy, are shown Fig.5a for different sample tilt angles. The dispersion branches for SPPs in the positive or negative $x$ directions (subscript signs in Fig.3 and Eqs.1,2) are indicated by circles and colored for the vacuum (blue) or $Si_3N_4$ (green) interfaces. The calculation nicely reproduces the experimentally observed electron DR of Fig.2b showing the inclined curves by the sample tilt. For comparison, the SPP DR of the corresponding sample tilt angles are overlaid in Fig. 2b. The slight mismatch can be attributed to the inaccuracy of the tilt angle and materials constant. The transferred momentum to the sample can also be calculated from Eq (2), as plotted in Fig.5b-e for different tilt conditions. By tilting the sample, $x$ component momentum $q_x^s$ is introduced to the sample (Fig. 5b,d), which causes the inclination of the apparent DR of the electron (Fig. 5a) .

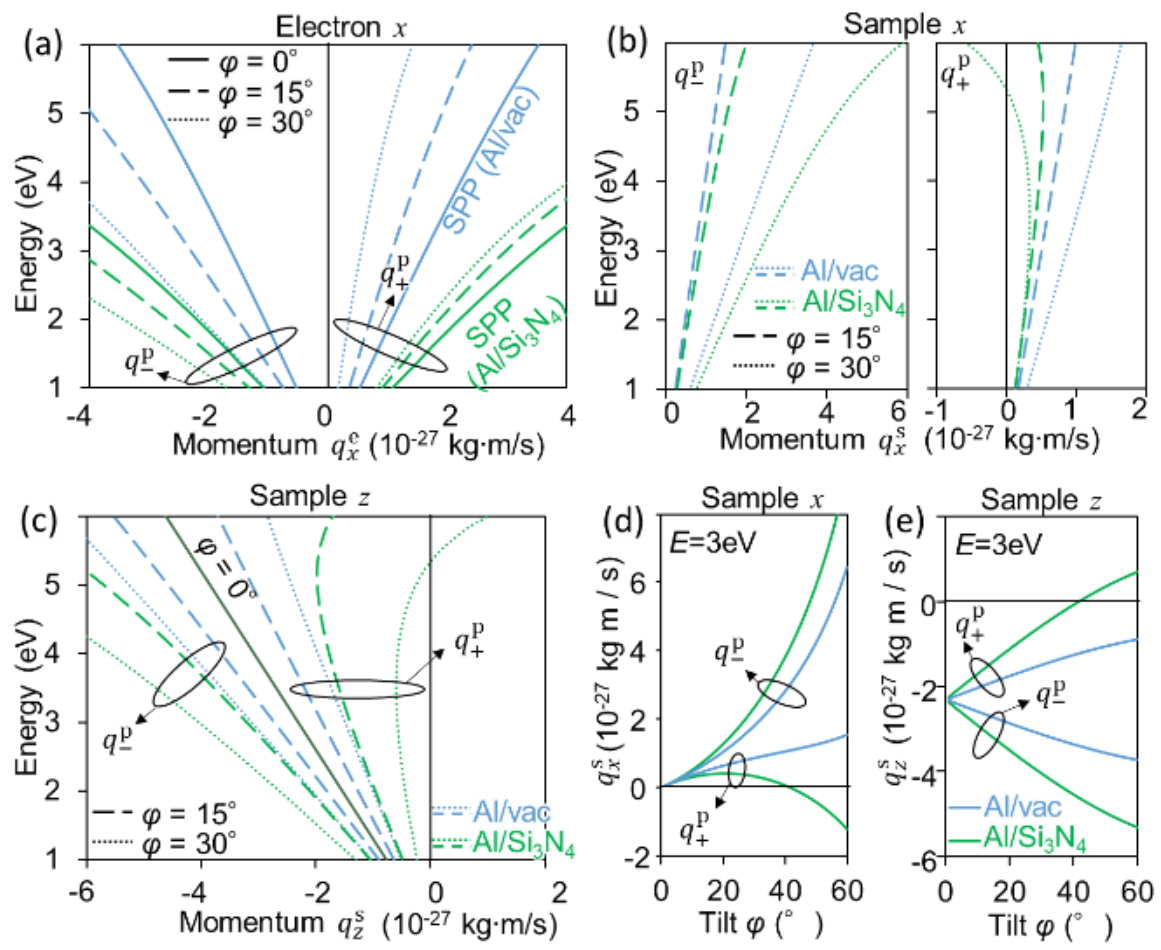

FIG 5. (a) Calculated in-plane electron DR with different tilt angles. (b,c) Transferred momentum to the sample in the (b) in-plane $x$ (plots separated for two scattering branches) and (c) out-of-plane $z$ directions. (d,e) Transferred momentum plotted as functions of the sample tilt angle.

The $z$ momentum introduced to the sample is downward ($q_z^s < 0$ in Fig.5c,e) in most conditions, which is natural since the fast electron is pushing down the sample. However, for the SPPs on $Si_3N_4$ with high angle tilts, the sample can be pushed upwards ($q_z^s > 0$) against the electron traveling direction, as a result of launching a SPP downwards (e.g. $\varphi = 30°$, $E > 5$ eV, Al/$Si_3N_4$, $q_+^p$ branch in Fig.5c, or $\varphi > 40°$ in Fig.5e). This condition coincides with the sample momentum sign change in the $x$ direction (Fig.5b,d). The high tilt angle results in SM indeed show dispersion branches in such counter-push conditions. [19] This may happen only when $q^p > \Delta q^e$, i.e. the optical mode cylinder radius is larger than the electron momentum change. In contrast, free photons do not fulfill this condition. SPPs of high momenta, such as those at dielectric/metal interfaces, with high sample tilt angle is required to fulfill this condition such that the total momentum of the electron, sample, and optical mode is conserved.

To include the periodicity, as in the hole array, one can add the SPP cylinders at the reciprocal lattice points, which could involve free space photons when the photonic mode dispersion goes inside the light line. Photon conversion can be explained by introducing light sphere. (see SM for details [19]) In such cases, where photons are emitted, electron-photon correlation measurements would allow accessing the sample momentum. [11,14,15]

In conclusion, we have demonstrated the modulation of the DR obtained in qEELS measurement by transferring the momentum to the sample through the sample tilt. Such momentum transfer to the sample is important to analyze the correlation, including entanglement, of the electron and electromagnetic mode. Although we here discussed the sample momentum transfer limited in the perpendicular direction, more degree of freedom should be taken into account for structured samples. A proper dimension design, such as periodicity, may allow investigating the entanglement involving samples. [19,27] For a small enough object, one could possibly monitor the sample momentum, where the kinetic energy of the sample is also involved. [28–30]

## ACKNOWLEDGMENTS

This work is supported by JST CREST JPMJCR25I3, FOREST JPMJFR213J, and JSPS Kakenhi 24H00400.

# Supplemental Material for

# Electron Recoil via Sample Momentum Transfer in Optical-Mode Excitation

Akira Yasuhara[1], Yamato Kirii[2], Takumi Sannomiya[2,*]

[1]JEOL Ltd., 3-1-2 Musashino, Akishima, Tokyo, 196-8558, Japan.
[2] Department of Materials Science and Engineering, School of Materials and Chemical Technologies, Institute of Science Tokyo, 4259 Nagatsuta, Midoriku, Yokohama, 226-8503, Japan

## S1. Mode analysis

The dispersion relation of the multilayer system is shown in Fig.S1a. For the silicon nitride film, a constant refractive index of 2.0 + 0.01i was used, which was confirmed by a separate optical transmission experiment together with the film thickness of 180 nm. We plotted the inverse of the determinant of the eigen value calculation matrix *X*, which expresses the series of boundary conditions. [1] For the single interface SPP dispersion, commonly used formulation of $k = \frac{\omega}{c}\sqrt{\frac{\varepsilon_1\varepsilon_2}{\varepsilon_1+\varepsilon_2}}$ is used. ($k$: wavenumber, $\omega$: angular frequency, $c$:light speed, $\varepsilon_1$:dielectric constant of the medium, $\varepsilon_2$: dielectric function of the metal.). Here the single interface SPPs well describe this system, meaning that the hybridization of SPPs on the vacuum and silicon nitride interfaces is not significant and that the top and bottom metal layers do not strongly couple through the silicon nitride layer. It is noted that the light line of the dielectric slab gives some intensity in this plot. Although the dielectric modes with different orders appear, they are not as efficiently excited as SPPs by a fast electron; The dielectric modes require excitation through Cherenkov radiation (CR) or transition radiation (TR). However, considering the momentum matching and excitation efficiency of CR and TR especially at higher mode energies than 2 eV, [2] the dielectric mode excitations are an order of magnitude smaller compared to that of SPPs. Indeed, we could observe only SPPs in EELS as shown in Fig.2b.

We also inspected the band gap of the hexagonal hole array of this multilayer by finite element method eigenmode analysis using COMSOL Multiphysics. We extracted the eigenmodes corresponding to the SPPs on the vacuum and silicon nitride interfaces. The eigenmode analysis result in FigS1b shows band gap formation of ~ 0.2 eV at the M point while that at the Γ point is smaller. These are not resolvable with a non-monochromated electron beam used in this study. We also note that the mode branch inside the light line (folded one), which are emissive, are not efficiently excited by free electrons (especially in the flat configuration) compared to the non-folded mode outside the light line because emissive modes have less electric field components along the electron beam path (Ez) than the propagating ones.

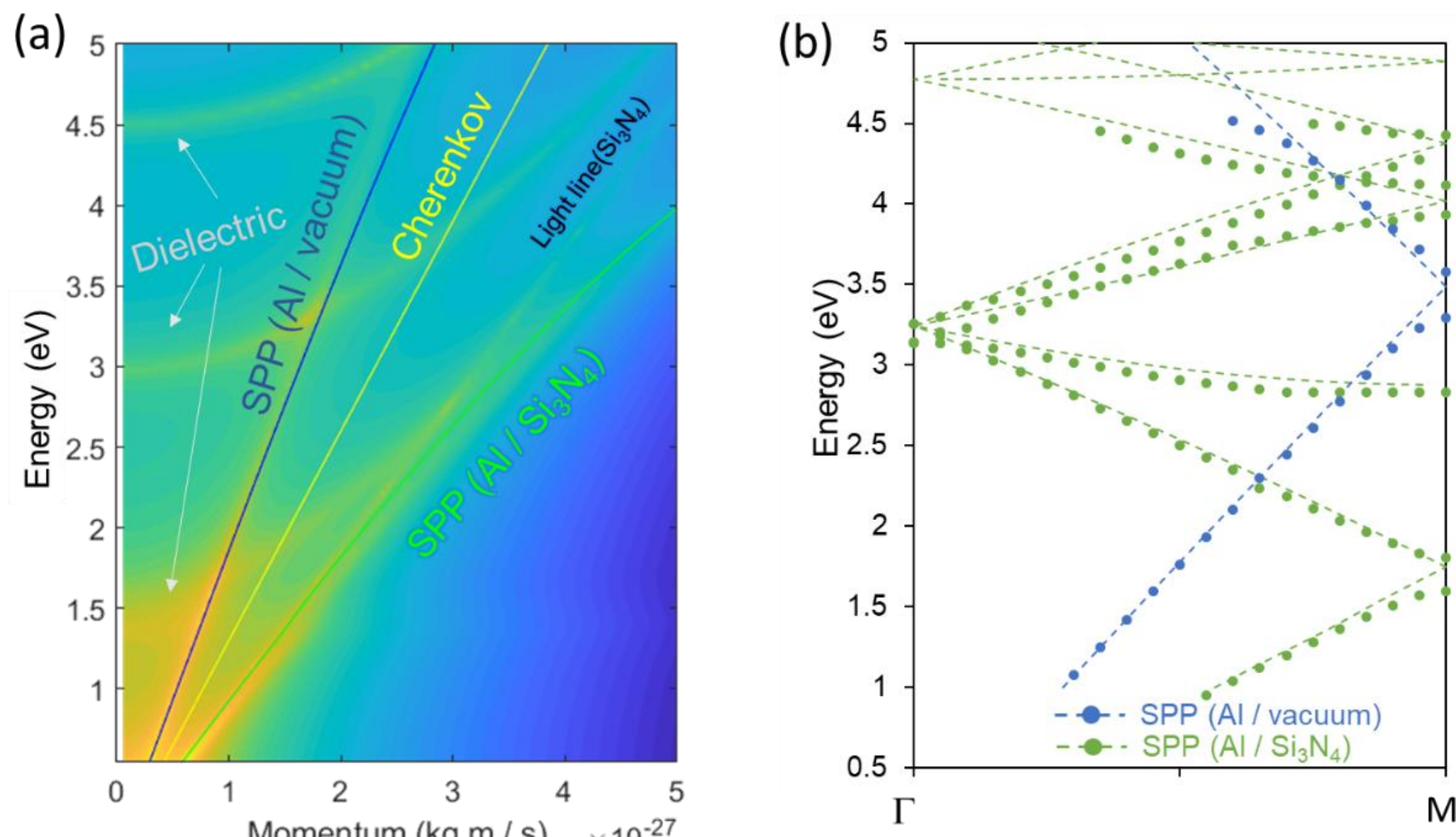

Fig.S1. (a) Dispersion plot of the experimental layer system (vacuum / Al 20nm / $Si_3N_4$ 180nm / Al 20nn / vacuum). The inverse of the determinant of the eigenvalue calculation matrix *X* is plotted. The overlaid SPP lines (blue and green) are the calculated for single interfaces. The yellow line shows Cherenkov radiation in $Si_3N_4$ for a 200 keV electron. The additional peaks correspond to the dielectric modes in $Si_3N_4$, which are not efficiently excited by free electrons. (b) Eigenmode analysis of the hexagonal hole array of this layer along the Γ-M direction corresponding to the experiment by finite element method (COMSOL) simulation. The dashed lines show the dispersions of an empty lattice approximation.

## S2. Reciprocal lattice

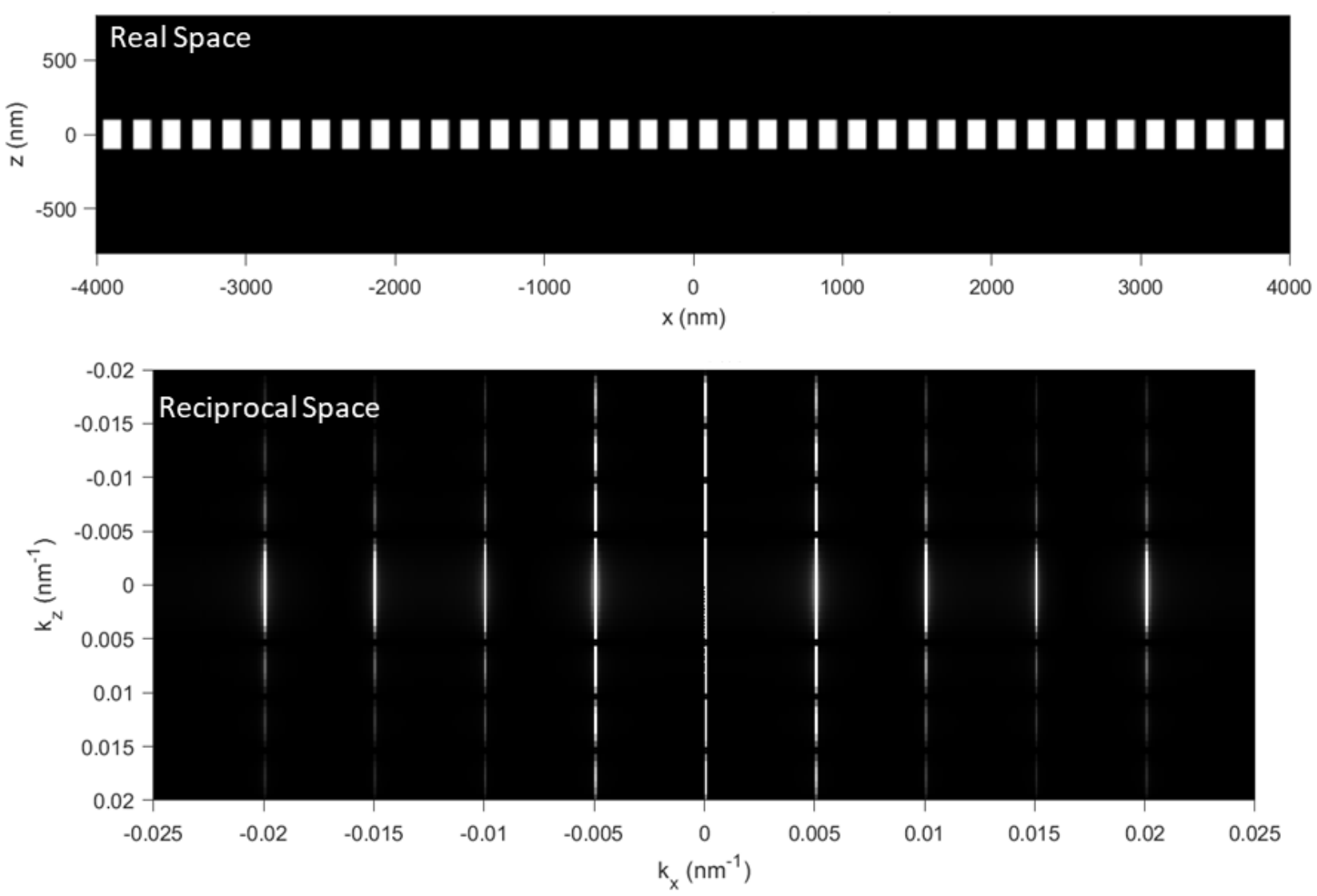

Fig.S2. Model calculation for the nanohole array structure cross section in the real space (top) and its reciprocal lattice (bottom) corresponding to the momentum space. The fringe in the z direction appears due to the thickness.

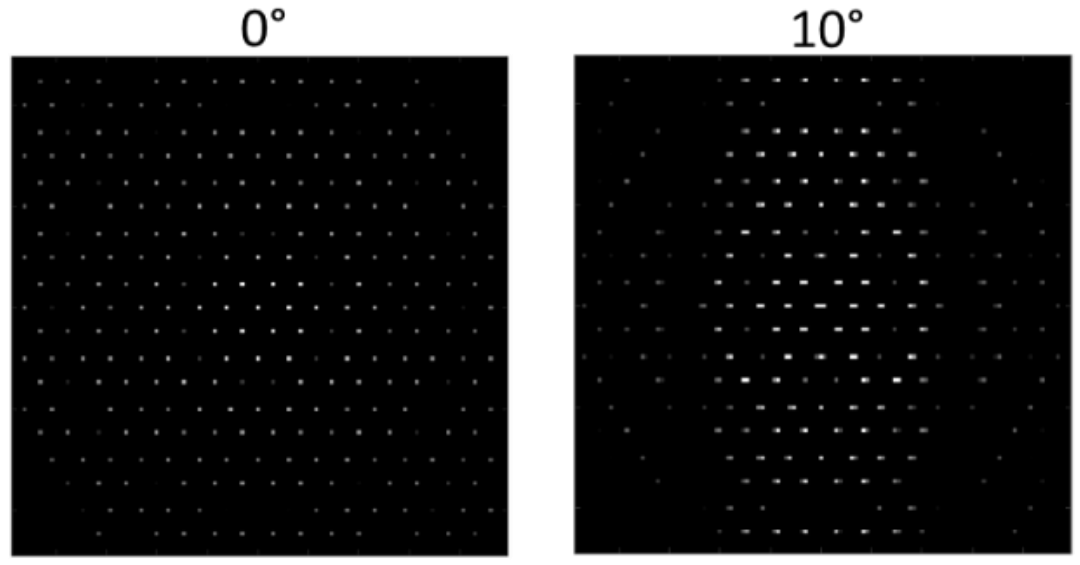

Fig.S3. Numerically simulated diffraction patterns of the hexagonal hole array film for $\varphi = 0$ and 10°. For the numerical load and for the simplicity, the simulation is performed only kinematically with a flat Ewald sphere, where no excitation errors nor dynamic effect is considered. Therefore, the asymmetry of the diffraction intensities on the left and right side, breaking the Friedel's law, is not reproduced.

## S3. 3D momentum space

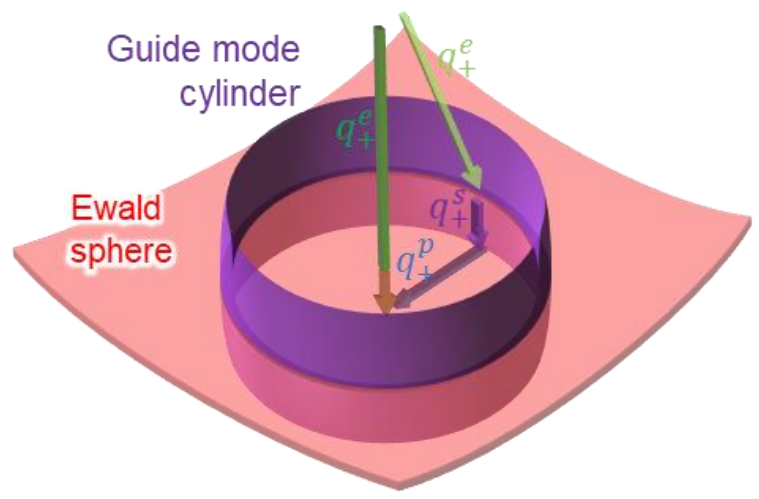

Fig.S4. 3D representation of the guide mode cylinder crossing the Ewald sphere.

## S4. High sample tilt results

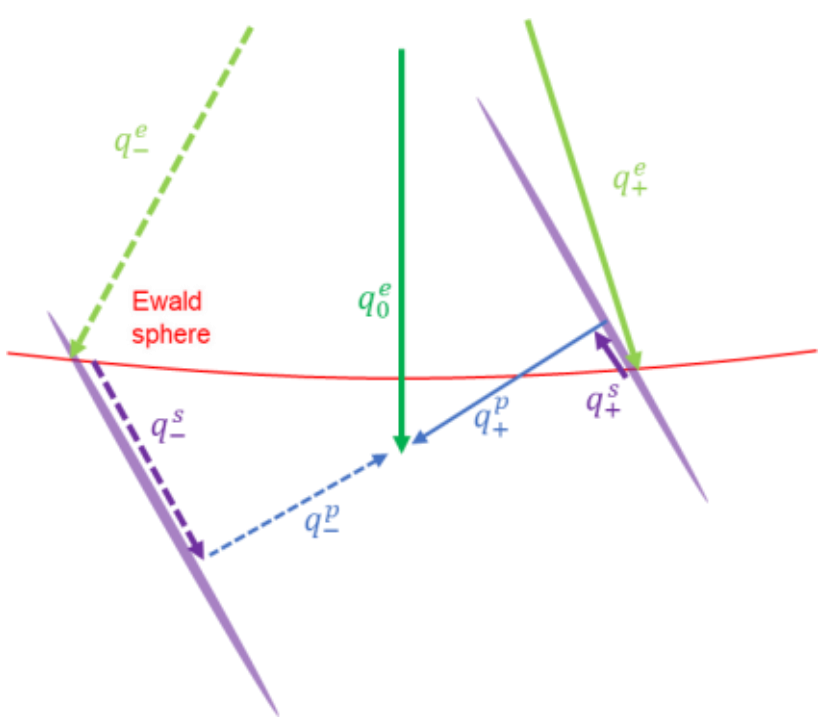

Fig.S5. (a) Illustration of momentum space with a large guide mode momentum $q^{\mathrm{p}}$ and high sample tilt angle $\varphi$. The momentum direction of the sample $q_+^{\mathrm{s}}$ may flip.

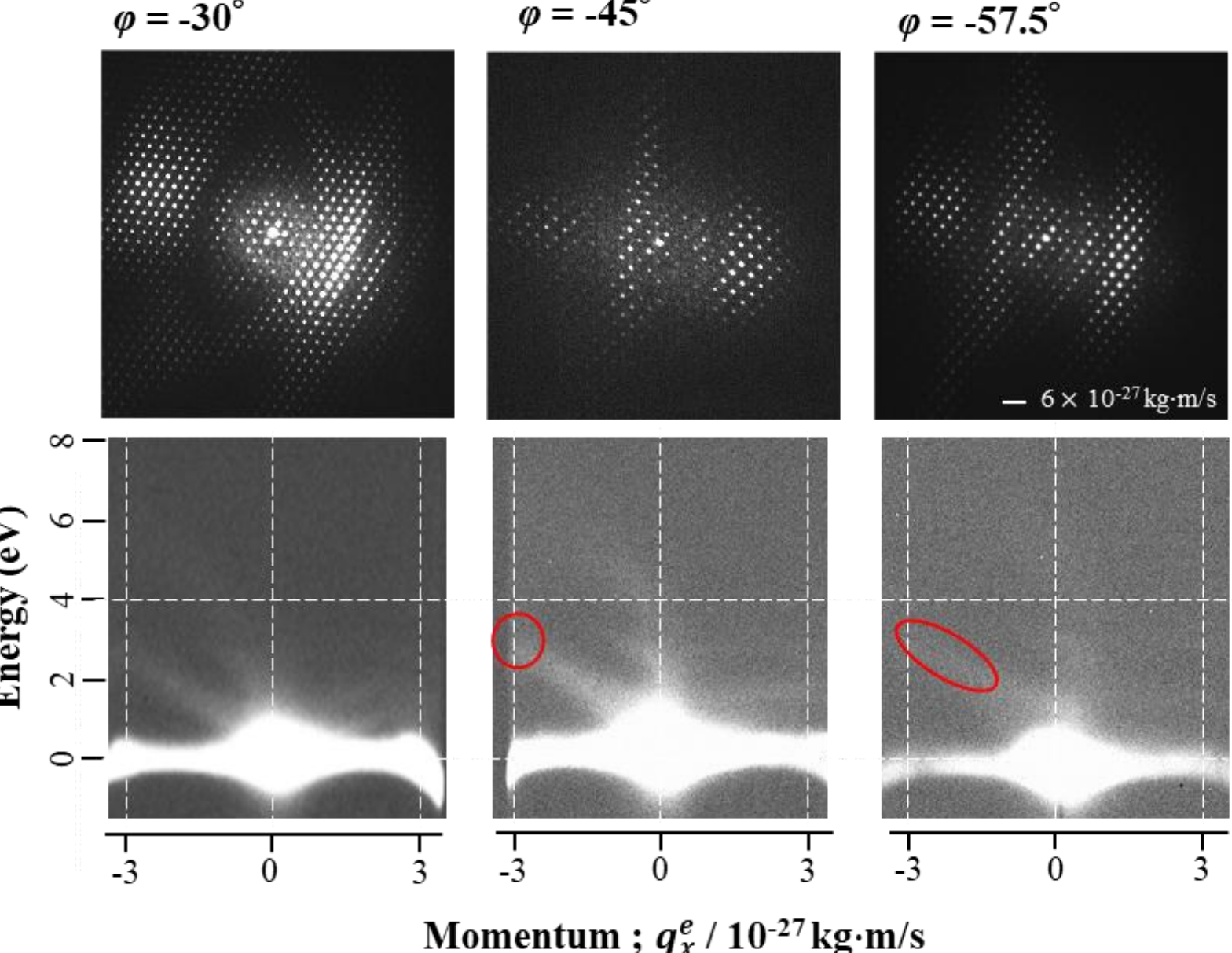

Fig.S6. qEELS measurement at high sample tilt angles. The dispersion becomes more difficult to recognize because of the increase of the effective sample thickness. The intersection of the lattice modes are visible close to the zero momentum and at lower energies, where the signal is stronger. The sample tilt alignment is not perfect because of the single-tilt holder for high angles up to 60°. The "upward push" conditions are marked by red circles.

## S5. Discussion on periodic structure and emissive mode

When the sample is periodic, some portion of the dispersion line of the optical mode (or SPP in this case) can appear inside the light line (i.e. photons can be generated), as described in Fig. S7a. We here consider the in-plane optical mode energy (thus, electron loss energy), as shown as the orange dotted line in Fig. S7a. For a given electron scattering, multiple scenarios, which cannot be distinguished only by the electron measurement, are probable: i) A photon is emitted because the green circle point is inside the light line (Fig. S7b). Here, a photon can be emitted upwards or downwards if the sample film allows photon emission on both sides. ii) Only in-plane optical mode is excited. (Fig. S7c). In this case, the momentum transfer G due to periodicity can be introduced as $q^G$ , which can be through the sample or directly in the photonics mode (not distinguishable). The former process with photon emission becomes more deterministic e.g. by an electron-photon correlation measurement although the latter process might be superposed. In the pure photon scinario, the photon and the sample momenta are entangled.

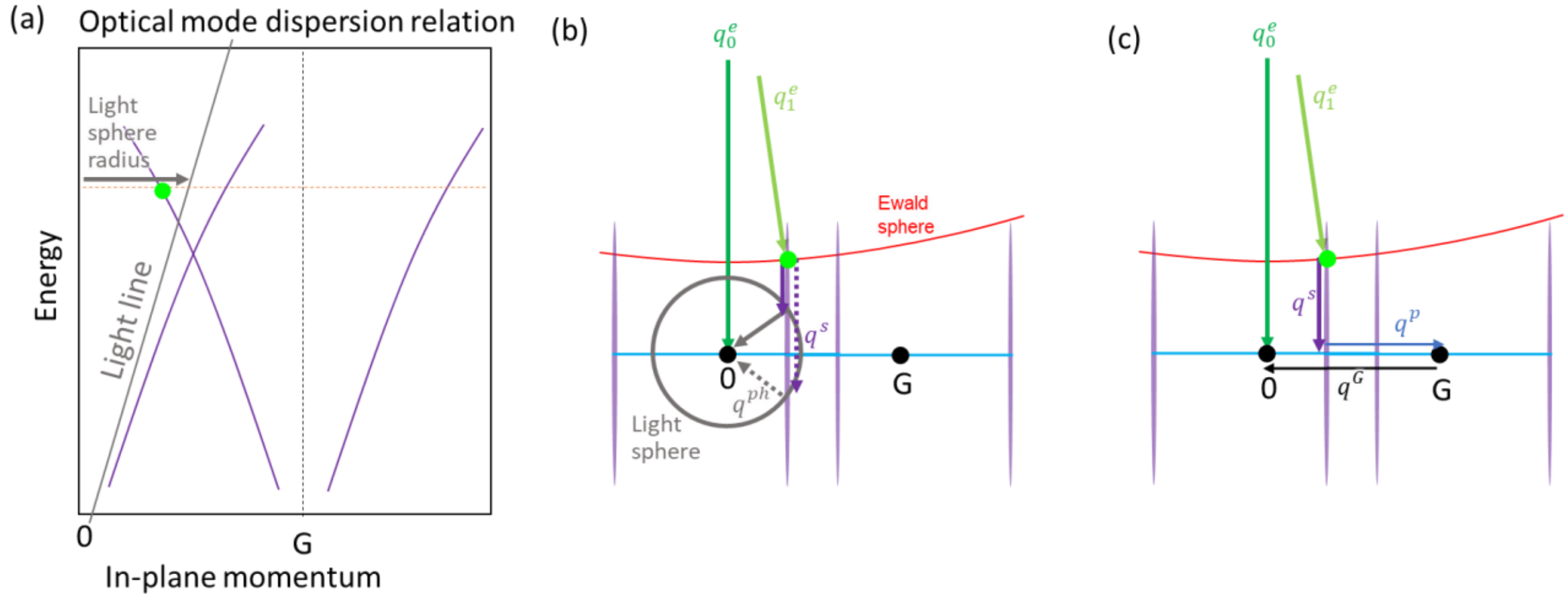

Fig. S7. Illustration of the momentum conservation with a lattice structure having a lattice point G. (a) Illustration of a SPP dispersion relation, where we consider the energy of the orange dotted line and the electron scattering corresponding to the green circle point. (b) Scattering process where a photon is emitted. (c) Scattering process without photon emission, i.e. only with optical mode excitation.

For the photon emission case, we see that the light sphere does not intersect the Ewald sphere, which can be understood as a free electron not emitting light in vacuum; For a free electron of energy $E_0$, the total momentum change $\Delta q^{\mathrm{e}}$ by the energy loss $\Delta E$ can be expressed as $\Delta q^{\mathrm{e}} = \frac{\Delta E}{\beta c}$, where $\beta = \sqrt{1 - 1/\gamma^2}$ and $\gamma = \frac{E_0 + mc^2}{mc^2}$. ($m$: electron mass, $c$: lightspeed) In contrast, the momentum of a free photon is $q^{\mathrm{ph}} = \frac{\Delta E}{c} < \Delta q^{\mathrm{e}}$. Thus the vacuum light sphere radius $q^{\mathrm{ph}}$ does not reach the Ewald sphere, as shown in Fig. S7b. However, the light sphere in a high index medium can be large enough such that the medium light sphere can intersect the Ewald sphere, that is Cherenkov radiation. (Fig. S8)

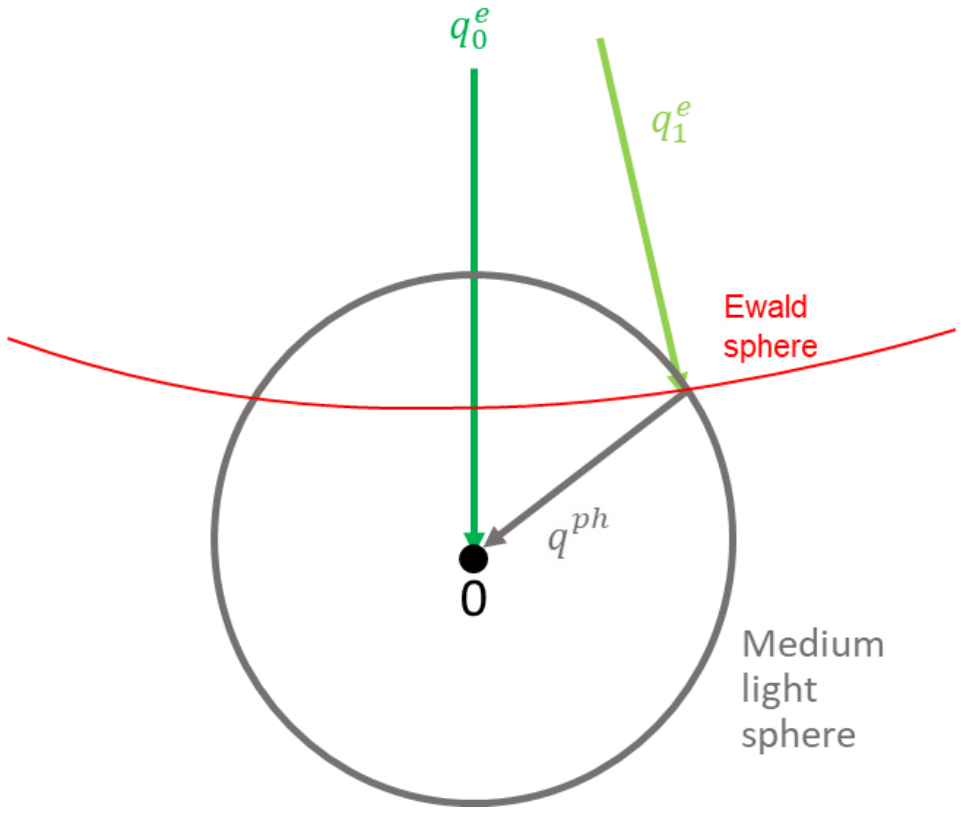

Fig. S8. Cherenkov radiation in the momentum space.

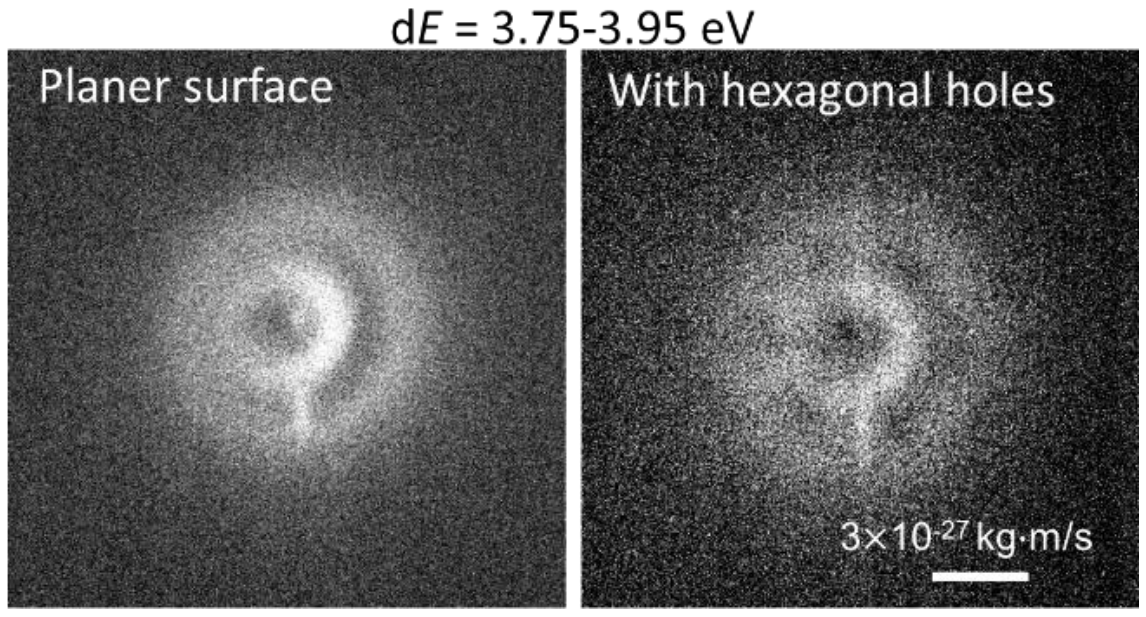

Fig. S9. Energy-filtered diffraction patterns (left) on a planer surface and (right) with a hexagonal hole lattice. The sample tilt angle is set to 15° and the energy loss range to 3.75-3.95 eV. The dispersion overlap from the lattice points are slightly visible.